\documentclass[floatfix,prd,lengthcheck,showpacs,amssymb,amsmath,amsfonts,aps,altaffilletter,nofootinbib,nopreprintnumbers,showpacsm]{revtex4-1}

 \usepackage[utf8]{inputenc}
\usepackage{color}
\usepackage{graphicx}
\usepackage{tabularx}
\usepackage{float}
\usepackage{subfigure}
\usepackage{amsmath}
\usepackage{acronym}
\usepackage{multirow}
\usepackage{tabu}
\usepackage{tikz}
\usetikzlibrary{arrows,shapes,trees,decorations.pathreplacing}
\usepackage{import}
\usepackage{svn-multi}
\usepackage{hyperref}
\usepackage{xspace}
\hypersetup{
    colorlinks=true,
    linkcolor=blue,
    filecolor=magenta,      
    urlcolor=cyan,
}

\newcommand{\result}[1]{{#1}}
\newcommand{\derek}[1]{\textsf{\color{purple}{[DD: #1]}}}
\newcommand{\laura}[1]{\textsf{\color{blue}{[LN: #1]}}}

\newcommand{\comment}[1]{}

\newcommand{\idq}{\texttt{iDQ}\xspace}
\newcommand{\pycbc}{\texttt{PyCBC}\xspace}

\newcolumntype{b}{>{\hsize=1.4\hsize}X}
\newcolumntype{s}{>{\hsize=.6\hsize}X}

\begin{document}

\title{Incorporating information from LIGO data quality streams into the PyCBC search for gravitational waves}
\author{Derek Davis$^1$,
        Max Trevor$^2$,
        Simone Mozzon$^3$,
        Laura K. Nuttall$^3$
        }

\address{$^1$ LIGO, California Institute of Technology, Pasadena, CA 91125, USA}
\address{$^2$ University of Maryland, College Park, MD 20742, USA}
\address{$^3$ University of Portsmouth, Portsmouth, PO1 3FX, UK}
\date{\today}

\begin{abstract}
We present a new method which accounts for changes in the properties of gravitational-wave detector noise over time in the PyCBC search for gravitational waves from compact binary coalescences. 
We use information from LIGO data quality streams that monitor the status of each detector and its environment to model changes in the rate of noise in each detector.
These data quality streams allow candidates identified in the data during periods of detector malfunctions to be more efficiently rejected as noise.
This method allows data from machine learning predictions of the detector state to be included as part of the PyCBC search, increasing the the total number of detectable gravitational-wave signals by up to \result{5\%}.
When both machine learning classifications and manually-generated flags are used to search data from LIGO-Virgo's third observing run, the total number of detectable gravitational-wave signals is increased by up to \result{20\%} compared to not using any data quality streams.
We also show how this method is flexible enough to include information from large numbers of additional arbitrary data streams that may be able to further increase the sensitivity of the search.
\end{abstract}

\maketitle

\section{Introduction}
\label{s:intro}

In the years since the first detection of gravitational waves by LIGO-Virgo~\cite{TheLIGOScientific:2014jea,VIRGO:2014yos,GW150914_paper}, 
the rate of detection has grown by over an order of magnitude~\cite{GWTC-1, GWTC-2, GWTC-3, GWTC-2.1}.
However, identifying gravitational waves in the collected data still requires 
the use of analysis pipelines that carefully look through the data. 
To date, all events detected by LIGO have been identified by at least one pipeline 
that uses matched filtering~\cite{Wiener:1949,Cutler:1992tc,Allen:2005fk}. 
A wide variety of matched filter pipelines have been developed
to analyze LIGO data~\cite{Messick:2016aqy,Aubin:2020goo,Chu:2020pjv,Venumadhav:2019lyq}. 
One such pipeline that has been in use since the first detection
of gravitational waves utilizes the \pycbc software suite~\cite{pycbc_release}. 
We refer to the offline search configuration of this pipeline~\cite{Usman:2015kfa,Nitz:2017svb,Davies:2020tsx} as \pycbc.

\comment{
Gravitational waves are observed with laser interferometers
designed to measure the strain caused by the 
passing through the detectors~\cite{TheLIGOScientific:2014jea}. \laura{ Not keen on the wording here. I think we should say wither GWs have been observed or that ifos are one way in which to observed GWs. Always wary about the other areas of GW astronomy.}
While this single strain time series encodes all of the available gravitational-wave information
and is sufficient to detect gravitational waves~\cite{TheLIGOScientific:2017qsa,Abbott:2017gyy},
\derek{We use matched filtering to search the data for GW.} 
}

The \pycbc search for gravitational waves from compact binary coalescenses (CBCs), is one of the main matched filter searches used to identify signals in LIGO data. 
\pycbc has been used to identified the vast majority of gravitational-wave signals to date~\cite{GWTC-1, 2-OGC, GWTC-2, 3-OGC, GWTC-3, 4-OGC, GWTC-2.1}.
As a matched filter search pipeline, \pycbc uses templates based on numerical models of gravitational-wave signals~\cite{Buonanno:2009zt,Bohe:2016gbl} to identify similar features in gravitational-wave interferometer strain data. 
Peaks in the matched filter signal-to-noise ratio (SNR) time series (referred to as ``triggers'') are found from these templates, and coincident sets of these triggers are assigned a ranking statistic that captures how likely it is that each trigger is a candidate gravitational-wave signal. 
The significance of these candidates is then estimated by simulated large amounts of background data by shifting the time stamp of triggers in one detector more than the gravitational-wave travel time between each site~\cite{LIGOS2iul}. 

One of the main challenges to detecting gravitational waves with matched
filter searches is the presence of non-Gaussian noise artifacts in the data. 
These artifacts are bursts of excess power that are referred to as ``glitches.''
Glitches are problematic for searches for gravitational waves  
as they can mimic or mask some features of astrophysical signals. 
It is also known that specific glitches can impact the measured search background~\cite{Cabero:2019orq,Davis:2020nyf}. 
As glitches are known to not be astrophysical in origin, 
it is imperative that \pycbc does not mistake a glitch for a real signal. 
Numerous features in \pycbc are designed to better differentiate
glitches from gravitational-wave signals using the gravitational-wave strain data alone. 
However, any additional information that can better help \pycbc
differentiate signals from glitches will improve the ability 
of \pycbc to identify gravitational-wave events. 

In recent observing runs, hundreds of thousands
of additional data streams beyond the gravitational-wave strain data were recorded
at each LIGO observatory; 
these data streams were used to both operate the detector
and monitor the detector environment~\cite{aLIGO:2020wna,AdvLIGO:2021oxw}.
These additional data streams, referred to as ``auxiliary data,'' 
can also be used to identify the source of glitches in LIGO data or
predict the presence of a glitch in the strain data. 
Auxiliary data has been shown to be beneficial for use in gravitational-wave
analyses to 
reject potential candidates due to noise~\cite{TheLIGOScientific:2017lwt}, 
subtract contributions of persistent noise from the strain time series~\cite{Driggers:2018gii, Davis:2019}, 
and validate the astrophysical origin of observed gravitational-wave events~\cite{LIGO:2021ppb}. 
Information from auxiliary data streams is generally re-packaged into more informative 
``data quality products'' that 
are used by \pycbc. 

At present, there are two main types of data quality products that are used by searches for gravitational waves from compact binary coalescences. 
The first, referred to as 
``data quality flags''~\cite{GW150914_detchar,LIGO:2021ppb}, 
are lists of time segments that are likely to contain glitches
based on information from specific auxiliary data streams. 
The most common way that a data quality flag is developed is by setting a threshold on the band-limited root-mean-square of an auxiliary data stream that is known to be correlated with the presence of glitches in the gravitation-wave strain data. 
The set of times where this threshold are exceeded are used as the data quality flag segments. 
Data quality flags 
have been used in most searches for gravitational waves to date~\cite{GWTC-1,GWTC-2,3-OGC,Venumadhav:2019lyq}. 
\pycbc uses data quality flags to reject candidates that occur
during the data quality flag segments or remove the time segment from the analysis 
completely.
An additional product, 
the \idq time series~\cite{Essick:2020qpo}, 
was introduced in LIGO's third observing run (O3). 
The \idq time series is a machine-learning based data quality product
that uses auxiliary data information to predict the likelihood
of a glitch being present in the strain data.
Methods incorporating \idq information have also been used in a recent search of LIGO data~\cite{Godwin:2020weu}. 
The process of developing and finalizing these curated data quality products
has taken multiple months in previous observing runs~\cite{M1000066}, 
increasing the total amount of time required to 
complete end-to-end analyses of LIGO data. 

In this work, we introduce a new method to incorporate information from these
data quality products into the \pycbc search for gravitational waves. 
This method is designed to use information from data quality products while
evaluating the significance of a given gravitational wave candidate, 
rather than simply rejecting candidates that occur during data quality flag segments.
This is accomplished by using data quality information as a part of the statistic used
to rank candidates in \pycbc.
We will also demonstrate how this method is generic enough to allow any data stream 
to be incorporated into the search pipeline, including the \idq time series, data quality flags, and the large number of auxiliary data streams that are recorded at each site.
While the methods described in this text are generalizable to the analysis of data from all ground-based gravitational interferometers,
such as Virgo~\cite{VIRGO:2014yos} or KAGRA~\cite{KAGRA:2018plz},
we exclusively work with data from the two LIGO detectors, 
LIGO Hanford and LIGO Livingston, in this work. 

We find that use of this new method increases the number of detectable
gravitational-waves
in a variety of different applications. 
In general, the increase in sensitivity when using these data quality streams was higher when using stricter thresholds for detection and when considering signals with higher masses.
Using data quality flags as part of the \pycbc search statistic rather than to reject candidates increases the search sensitivity to \result{10\%} for the highest masses and the strictest detection threshold considered in this work. 
Including \idq information via this method also increases the sensitivity by a further \result{5\%}. 
We also show that directly using information from auxiliary data streams that monitor seismic noise can improve the sensitivity of \pycbc by up to \result{5\%}. 
Finally, we test all auxiliary data streams that are currently publicly available~\cite{auxdata} and identify \result{10} streams that show significant correlations. 
If used in an end-to-end search for gravitational waves,
use of additional auxiliary data is also likely to increase the search sensitivity. 

This work is organized as follows. 
In the remainder of this section, we outline the current methods used in the \pycbc
search for compact binaries~\cite{Usman:2015kfa,Nitz:2017svb,Davies:2020tsx}, 
with emphasis on how the properties of the detector noise are modelled in the search. 
We also discuss some of the current products that are produced by the LIGO collaboration to track the data quality. 
We then explain, in Section~\ref{sec:model}, our proposed improvement to the noise model 
in \pycbc, and how this improved model can be used in a variety of cases. 
We demonstrate the benefits of our improved model for multiple applications in Section~\ref{sec:app}.
Finally, we discuss how this improved model will benefit future searches for gravitational waves in Section~\ref{sec:conc}.

\subsection{Identifying Gravitational-wave Signals}
\label{sec:extract}

The \pycbc search for compact binaries~\cite{Usman:2015kfa,Nitz:2017svb,Davies:2020tsx} 
identifies gravitational-wave events using matched filtering with gravitational waveforms predicted by general relativity~\cite{Buonanno:2009zt,Bohe:2016gbl}. 
The SNR for a matched filter with a specific waveform template $h$ is~\cite{Allen:2005fk}

\begin{equation}
 \rho^2 (t) \equiv \frac{\|\left\langle s | h \right\rangle\|^2}{\left\langle h | h \right\rangle} 
 \text{ ,}
\end{equation}
\label{eq:snr}
where the inner product, $\langle | \rangle$, is defined as

\begin{equation}
 \langle a|b\rangle (t) = 4 \text{Re} \int^\infty_0 \frac{\tilde{a}(f)\tilde{b}^*(f)}{S_n(f)} 
 e^{2 \pi i t f} df 
 \text{ ,}
\end{equation}
with $s$ the strain data,
$h$ the template, 
and $S_n(f)$ the estimated power spectral 
density for the time in question. 
This is equivalent to
cross-correlation in the frequency domain. 
Peaks in this SNR time series are labelled as triggers and correspond to potential 
signals in the data that are similar to the template. 

If data from gravitational-wave interferometers were purely stationary and Gaussian noise, 
the matched filter SNR would be sufficient to identify signals in the data. 
However, variations in the properties of the noise, both over short and long periods, 
complicate the problem. 
To account for the non-idealized features in the data, 
a ``ranking statistic'' is created that includes additional information about the data and expected signal properties
to better differentiate signals from noise. 
Ideally, an astrophysical signal should receive a high ranking statistic, 
while a noise fluctuation should receive a lower ranking statistic 
(often referred to as being `down-ranked''). 

The generic form of the ranking statistic is given by
the ratio of the signal and noise distributions~\cite{Davies:2020tsx} for a given set of parameters, $\vec{\kappa}$, 

\begin{equation}
    \Lambda_{opt}(\vec{\kappa}) = \eta_S \frac{\hat{r_S}(\vec{\kappa})}{r_N(\vec{\kappa})}
    \text{ ,}
\end{equation}
where $\eta_S$ is the overall rate of signals and  
$\hat{r_S}(\vec{\kappa})$ is the transfer function between the true rate of signals the the detectable rate,  
and $r_N(\vec{\kappa})$ 
is the rate of noise.

It is convenient to consider the ratio of these two distributions 
as the difference of their logarithms, 

\begin{equation}
    R(\vec{\kappa}) = \log{r_{S}(\vec{\kappa})}-\log{r_{N}(\vec{\kappa})}
    \text{ .}
\end{equation}

The parameters, $\vec{\kappa}$, that are used in both the \pycbc signal and noise models
encode details about the physical properties of the triggers and how well the measured data matches
that expected of an astrophysical signal. 
In addition to the matched filter SNR, 
numerous signal consistency tests are included~\cite{Allen:2005fk,Nitz:2017lco}
to measure how well a candidate trigger matches the expected signal morphology.
The signal model is based on the expected distribution of these parameters for astrophysical signals. 
These parameters are then used to calculate a single reweighted SNR, $\hat{\rho}$, 
that quantifies how well the data matches a real signal in each detector. 
This value is generally referred to as the 
``single-detector statistic.''
Additional parameters that corresponds to relationships between the data in multiple detectors are also used. 
Full details of the signal model are provided in~\cite{Davies:2020tsx}.
We will outline the noise model in additional depth for convenience.

The \pycbc noise model is based on fitting the 
distribution of triggers in the data to an exponential decay function. 
The rate of noise triggers for a given template in a particular detector, $\vec{\theta}$, is fit to an exponential given by

\begin{equation}
    r_N(\hat{\rho};\vec{\theta}) =
    \mu(\vec{\theta}) \alpha(\vec{\theta})
    \exp{\left[ -\alpha(\vec{\theta}) (\hat{\rho}-\hat{\rho}_{th}) \right]}
    \text{ .}
\end{equation}

For a given template, $\vec{\theta}$, the term $\mu(\vec{\theta})$ is the number of triggers above threshold and $\alpha(\vec{\theta})$ is the exponential decay rate with respect to $\hat{\rho}$.
This $\hat{\rho}$ is the same single-detector ranking statistic that is used in the signal model.
Only triggers with $\hat{\rho} > \hat{\rho}_{th}$ are considered in this fit.

In this time-independent \pycbc noise model, a number of approximations are already used to calculate $\mu(\vec{\theta})$ and $\alpha(\vec{\theta})$~\cite{Nitz:2017svb}.
First, a maximum likelihood fit of this noise model is performed for for each detector and each template in the search individually.
However, there are not enough triggers identified per template to to accurately measure both parameters for each template.
Kernel smoothing is used to reduce noise in the the measured values of $\mu$ and $\alpha$ with respect to the duration of each template, $\tau(\vec{\theta})$.
Hence the noise model used in the search is $r_N(\hat{\rho};\vec{\theta}) \approx r_N(\hat{\rho};\tau(\vec{\theta}))$. 

This model of the noise does not include any time dependence, 
meaning that
this fit assumes a single distribution is valid for each template 
during the entire analysis period. 
To account for variation in the properties of the noise with respect to time, 
it is typical for the \pycbc search to be run separately over relatively short chunks of data
(typically 5 days). 
However, it is known~\cite{GW150914_detchar,LIGO:2021ppb} that gravitational-wave interferometer data contains short term fluctuations
on both the hour- and second-scale.
While some techniques have been developed to account for these fluctuations~\cite{Mozzon:2020gwa,Zackay:2019kkv}, 
they do not introduce explicit time-dependence into the noise model itself.

\subsection{LIGO Data Quality Information}
\label{sec:dq}

At each LIGO observatory, hundreds of thousands of data streams are recorded
during an observing run to control and monitor the detectors~\cite{AdvLIGO:2021oxw}. 
A subset of these data streams have been found to be highly correlated with periods of
excess noise in LIGO strain data.
For example, 
ground motion that introduces additional motion 
of the test masses and increases the chances of scattered light is well monitored by seismic sensors.
However, when this information is used to support the astrophysical analyses,
this data is first curated into data quality products. 
These data quality products combine multiple data streams into a 
single data product that is simpler for astrophysical analyses to utilize. 
Similar data quality products are produced for 
other gravitational-wave observatories~\cite{virgo_detchar,LIGOScientific:2022myk}. 

One example of a data quality product are  data quality 
flags~\cite{GW150914_detchar,TheLIGOScientific:2017lwt,LIGO:2021ppb}. 
Data quality flags are binary data streams sampled at 1~Hz that have multiple categories
to indicate the different severity of noise likely to be present in the detector. 
These flags are used by \pycbc to remove times from an analysis or
veto triggers identified during a data quality flag.
Other searches~\cite{Sachdev:2019vvd,Aubin:2020goo} instead use these flags to replace the data with zeroes during flagged times.

Another data quality product that has been used in analyses is 
the \idq time series~\cite{Essick:2020qpo}.
This product is based on a machine-learning algorithm that uses a large 
number of auxiliary data streams to predict
the likelihood of a glitch being present in the detector strain data at a given time. 
In O3, \idq was a single time series sampled at 128 Hz. 
One key difference compared to data quality flags is that \idq is not a binary data stream, 
and instead assigns a likelihood to each sample based on the probability that the strain data contains a glitch.
Methods to incorporate \idq into a different pipeline used to search for gravitational waves from compact binaries, 
GstLAL~\cite{Messick:2016aqy}, 
were recently developed~\cite{Godwin:2020weu}. 
This method directly used the \idq likelihood as
an additional term in the 
ranking statistic of the GstLAL.
Additional details on GstLAL can be found in~\cite{Sachdev:2019vvd,Hanna:2019ezx,Cannon:2020qnf}. 
Comparisons between the methods introduced in this work and those currently implemented in GstLAL are discussed in Section~\ref{sec:nonbinary}.

As use of curated data quality products has
been consistently shown to increase the sensitivity 
of searches for gravitational waves~\cite{TheLIGOScientific:2017lwt,Davis:2020nyf,Godwin:2020weu}, 
it is also likely that 
the auxiliary data used to generate these products can also benefit gravitational-wave searches. 
Furthermore, it is possible some useful information from the auxiliary data is discarded when curated data products are generated. 
Use of the raw auxiliary data is also attractive as 
it would reduce the time to complete an end-to-end analysis of LIGO data by no longer requiring time to generate data quality products. 
However,
the large number of different data streams with disparate properties has made it difficult to develop
generic methods to incorporate this data.

\section{Improved noise model}
\label{sec:model}

The changing state of the detectors means that the \pycbc background will also change with respect to time. 
Therefore a more complete description of the noise model that accounts for this time-dependence should be given as 

\begin{equation}
    r_N(\hat{\rho};\vec{\theta};t) =
    \mu(\vec{\theta};t) \alpha(\vec{\theta};t)
    \exp{\left[ -\alpha(\vec{\theta};t) (\hat{\rho}-\hat{\rho}_{th}) \right]}
    \text{ ,}
\end{equation}
where $\mu(\vec{\theta};t)$ is the trigger density for a given template 
with respect to time and 
$\alpha(\vec{\theta};t)$ is the exponential decay rate of the background 
for a given template with respect to time.

Due to practical limitations, we only consider the time-dependence of the trigger density, $\mu$,
and ignore the time-dependence of the decay rate, $\alpha$.
The time-dependent variations that we hope to capture with this improved noise model
can occur over timescales of seconds. 
Although a large number of triggers are identified by \pycbc per analysis, 
the average number of triggers per second in recent analyses is only \result{10 triggers per second}, 
even though hundreds of thousands of templates are used to search the data. 
The modest rate of triggers, combined with the large number of templates considered, 
means that there is much less than 1 trigger per template per second. 
This is not a sufficient number of triggers to accurately measure second-scale variations
in the trigger rate. 
However, if we suitably bin the times and templates, this is a sufficient number of triggers to 
measure the time-dependence of the trigger rate. 
When using bins, we approximate the overall time-dependence of the trigger rate as the originally measured trigger rate multiplied by an additional bin-dependent factor.

With these approximations, the only change to the noise model is to $\mu(\vec{\theta;t})$, where the time-dependence is modeled as the product of a time-independent term, $\mu(\vec{\theta})$ and a time-dependent term $\delta(\vec{\theta};t)$.
Hence the new expression for trigger density is

\begin{equation}
    \mu(\vec{\theta};t) \approx \mu(\vec{\theta})\delta(\vec{\theta};t)
\end{equation}

for a given $\theta$, and time, $t$. 
The calculation of $\delta({\theta};t)$ is different depending on the type of data quality stream that is being considered.

This method makes no assumptions about the 
input data that is used as part of the noise model. 
In the case that the auxiliary data is non-informative (i.e. 
not correlated with times of high trigger density),
the method should identify that no excess in triggers
is measured, and no change to the noise model will be applied. 

\subsection{Binning the parameter space}
\label{sec:bins}

\begin{figure*}[tb]
  \centering
    \includegraphics[width=\textwidth]{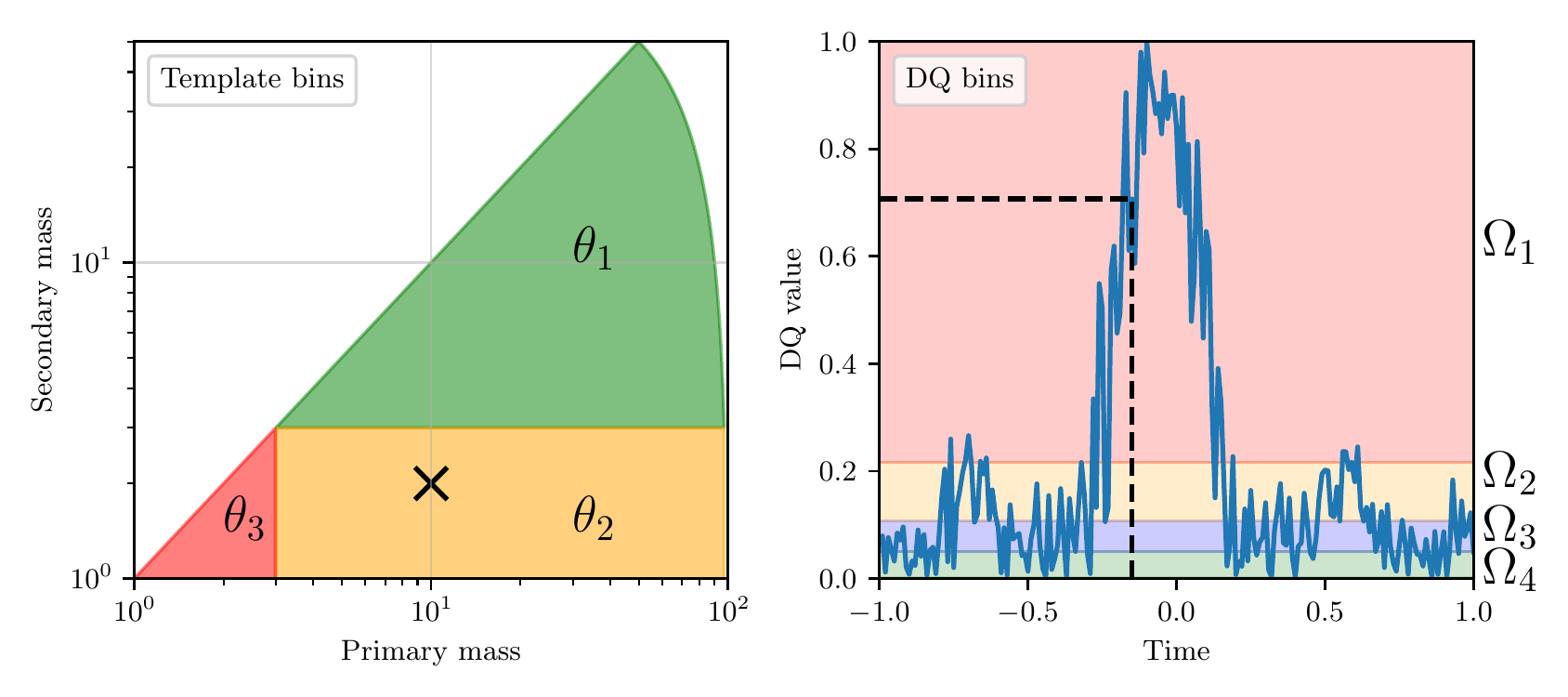}
\caption{
An example of how the template and data quality (DQ) bins are constructed and applied. 
\emph{Left:} A plot of the three different template bins corresponding to different parts of the template bank used in the search. 
In this case, the range of templates is characterized using the masses of the primary and secondary components of the simulated compact binary system template .  
An ``$\times$'' marks the template parameters of an example candidate is in the template bin $\theta_2$. 
\emph{Right:} A plot of an example data stream that is used to construct four different data quality bins. 
The dotted line marks the time of an example candidate. 
In this case, the example candidates is in
data quality bin $\Omega_1$.
Therefore the time-dependent term used in the \pycbc noise model for this candidate would be $\delta(\theta_2;\Omega_1)$.
}
\label{fig:bins}
\end{figure*}

In the current \pycbc noise model, the trigger rate, $\mu(\vec{\theta})$,  is calculated for each individual template.
In practice, it is not possible to also determine the time-dependent correction to the trigger rate, $\delta(\vec{\theta};t)$, for each individual template. 
This is due to the relatively low rate of triggers per template per second. 
In order to approximate the value of $\delta(\vec{\theta};t)$, 
we choose to group templates with similar duration 
into bins (denoted by $\theta_b$).
We also group times based on the value of the data quality stream, $\Omega(t)$, using additional bins (denoted by $\Omega_d$).
We then calculate a single value of $\delta({\theta_b};\Omega_d)$ for each combination of $\{\theta_b,\Omega_d\}$.
These data quality bins span the range of values that the data quality stream can take and a single time-dependent correction is calculated for all times in each data quality bin. 
If we have $N$ template bins and $M$ data-quality bins, this means we only need to estimate $ N \times M$
different corrections to the trigger rate.
An example of how these bins could be constructed with 
\result{3} template bins and 
\result{4} data quality bins, 
along with the parameters of an example trigger, 
is shown in Figure \ref{fig:bins}.

\comment{
First, we use the available data quality information 
to parameterize the time stream, as opposed to time itself
\laura{I don't understand what this means}. 
Using data quality information to consider longer segments of data increases the chance that the data considered in each of these chunks has similar background characteristics. 
The use of data quality information also reduces the complexity of the problem by consider the smaller number of possible data quality values rather than each moment of time individually.
\laura{actually I don't understand this paragraph.} 
}

We will label each template bin as $\theta_b$
and each data quality bin $\Omega_d$.
This means that the time-dependent correction 
the trigger rate in our noise model is defined as

\begin{equation}
    \mu(\vec{\theta};t) \approx \mu(\vec{\theta})\delta(\theta_b;\Omega_d)
    \text{ .}
\end{equation}

Care must be taken when deciding on the number of bins to use in an analysis. 
As the presence of a gravitational-wave signal will naturally cause more triggers to be observed, there is a risk that real signals will be down-ranked if the total number of triggers produced by a signal is a significant fraction of the total number of triggers in a single bin.
Conversely, if not enough bins are used, variations in time and across the template bank may not be captured. 
We found that having at least \result{50} triggers in each bin was sufficient to minimize the chance that a real signal would be artificially down-ranked.

In this work, we chose bin sizes such that the smallest bin contained at least this minimum number of triggers.
This resulted in the choice of 10 template bins
and either 
2 (the binary case where one bin is much smaller than the other) or 200 (the non-binary case where all bins are the same size) data quality bins. 
This means that either 20 or 2000 different
values of 
$\delta(\theta_b;\Omega_d)$
must be calculated for every data stream.

We construct our template bins based on template duration, with the goal of recording an equal number of triggers in each template bin.
A representative chunk of LIGO data from O3 was used to calculate the specific values of the bin edges used.
The bin edges are linearly spaced 
at $\{0, 10, 20 \ldots, 100\}$ percentile of the trigger template duration. 
After calculating the bin edges for this representative chunk of data, the same values of template duration were used as bin edges in all analyses. 
This is the default binning strategy used in this work. 
Two alternate binning strategies  were also investigated, but were found to result in a smaller sensitivity increase than our default strategy.
More details are given in Section~\ref{sec:app_bins}. 

\subsection{Binary data quality streams}
\label{sec:binary}

The simplest case we can consider is a binary data quality stream 
that only consists of 1s and 0s. Times where the data quality stream is 1 are often referred to as 
``active'' times, and times that the stream is 0 
are referred to as ``inactive'' times.
Data quality flags are one example of a binary data stream.
In this scenario, the time dependence of $\delta(\theta_b;\Omega_d)$
is also binary. 
We only have two data quality bins, 
labeled $\Omega_1$ and $\Omega_0$.
For times that the stream is active,
the time-dependent term of the noise model, 
$\delta(\theta_b;\Omega_1)$, is defined as

\begin{equation}
    \delta(\theta_b;\Omega_1) = 
    \frac{N_{b,1}}{T_{1}}
    \frac{T_{tot}}{N_{tot}}
\end{equation}
for a given template bin, $\theta_b$.
$N_{b,1}$ is the total number of triggers in template bin $b$
during times the data quality stream is active, 
while $N_{tot}$ is the total number of triggers in the analysis.
Similarly, $T_{1}$ is the total amount of time the data quality stream is active, 
while $T_{tot}$ is the total amount of time in the analysis.

If the binary data quality stream is correlated with periods
of high trigger density, 
then $\delta(\theta_b;\Omega_1) > 1$.
However, this is not guaranteed to be the case. 
If $\delta(\theta_b;\Omega_1) \leq 1$, we 
impose $\delta(\theta_b;\Omega_1) = 1$.
This is so that the data stream does not
increase the significance of a candidate. 
All times when the data quality stream is inactive 
are also fixed to $\delta(\theta_b;\Omega_0) = 1$.

\subsection{Non-binary data quality streams}
\label{sec:nonbinary}

We can also consider a data quality stream that takes an arbitrarily large number of values. 
Such data quality streams include the \idq time series
or auxiliary data.
In this case, we bin the data points into multiple data quality bins based on the value of each data point.
The total number of bins used with this method must be tuned for each analysis. 
In this work, we choose to use \result{200} data quality bins so that each bin contained a sufficient number of triggers to reduce the bias of individual astrophysical signals.  

The correction, $\delta(\theta_b;\Omega_d)$, for each data quality bin, $\Omega_d$ is calculated using the same
formula as the binary case.
Again similar to the binary case, we fix $\delta(\theta_b;\Omega_d) \geq 1$.
Times where the data quality stream is not defined are still still used to calculate the total time and total number of triggers, but these triggers during these times are not reranked using this method. 

We can also compare this correction to the model suggested for use with the \idq time series in~\cite{Godwin:2020weu}. 
There are two main differences between our model and the model from~\cite{Godwin:2020weu}.
Firstly, in our model the correction to the total trigger density is directly computed for each combination of 
template bin and data quality bin.
This ensures that an accurate correction is applied for any data quality stream.
Compared to the analytic model designed for use with the \idq time series described in~\cite{Godwin:2020weu},
there is a reduced risk of reranking candidates by too much or too little. 
Our method also does not impose a maximum correction to the trigger density as was done 
in~\cite{Godwin:2020weu}.
While this does introduce a risk of an arbitrarily high correction being applied, 
such a case would not occur unless there was indeed a strong correlation between the data quality 
stream and the \pycbc triggers, 
implying that the trigger is unlikely to come from an astrophysical signal.

\subsection{Multiple data quality streams}
\label{sec:multi}

For the time-dependent correction associated with two different data quality streams, $\delta_{n}$ and $\delta_{m}$, we define the joint time-dependent correction, $\delta_{nm}$, as

\begin{equation}
    \delta_{nm}(\theta_b;\Omega_d) = \max{\left( \delta_{n}(\theta_b;\Omega_d),\delta_{m}(\theta_b;\Omega_d) \right)}
    \text{ .}
\end{equation}

This conservative choice ensures that if more beneficial data quality
information is available, the relevant triggers will be down-ranked by a larger amount. 
The choice to downrank candidates by the largest time-dependent correction may lead to some triggers being down-ranked more or less than would be optimal. 
For example, this choice ignores any correlations between the two data quality streams.
However, as both astrophysical signals and triggers caused by noise are down-ranked the same amount, 
we do not expect this to decrease the sensitivity of the search as compared to not using any data quality streams. 

\section{Applications}
\label{sec:app}

One of the significant benefits of this method of incorporating 
data quality streams into the \pycbc search is
its versatility in a variety of applications. 
In this section, we will demonstrate a number of use cases
for this method and investigate
how incorporating each data quality stream increases
the detection rate of gravitational-wave signals
by the \pycbc search. 
In all cases,
we find evidence that incorporating these data quality 
streams can increase the number of detectable gravitational waves. 

The O3 strain data used in this section 
from both the LIGO Hanford and LIGO Livingston 
detectors is available 
from the Gravitational Wave Open Science Center (GWOSC)~\cite{LIGOScientific:2019lzm}. 
Although most auxiliary data recorded by LIGO 
is not yet publicly available, 
there has been a release of auxiliary data around one event~\cite{auxdata}
and a small number of data quality products that are released publicly alongside the strain data.
The majority of these analyses in this section demonstrate how data quality products not yet publicly released could be used to improve the sensitivity of the \pycbc search. 
The source of each data quality stream, either public or not public, is described in the relevant section. 

\subsection{Search configuration}
\label{sec:config}

The analyses presented in this section use data from 5
different analysis periods. 
These time periods correspond to the chunks of data analysed by the LIGO-Virgo collaborations
during O3. 
The start and end times of each chunk are listed in Table~\ref{tab:flag}.
We label each chunk by a number between 1 and 5. 
These chunks were chosen due to known data quality issues that
may impact the sensitivity of the \pycbc search. 

In all examples presented here, 
we use the ranking statistic introduced in this work and available as part of the \pycbc 
code repository found at~\cite{pycbc_release}.
We use a single-detector ranking statistic that includes the chi-squared test~\cite{Allen:2005fk}, the sine-Gaussian test~\cite{Nitz:2017lco}, and accounts for variation in the detector's power spectral density with time~\cite{Mozzon:2020gwa}.
We use the same template bank as was used in \pycbc analyses presented in GWTC-3~\cite{GWTC-3,Roy:2017qgg,Roy:2017oul,DalCanton:2017ala}.
Unless explicitly stated, triggers with a single-detector statistic above \result{6.5} are used to calculate the time-dependent correction to the \pycbc noise model. 
This threshold was tuned by hand to balance including a sufficient number of triggers to model the time-dependence and focusing on the tail of the non-Guassian distribution of triggers. 

We  compare the sensitivity of the search with and without incorporating data quality information by comparing the volume-time (VT) of the search in each case. 
This is done using a large number of simulated signals that are recovered by the \pycbc search pipeline. 
The distance at which a simulated signal can be detected is then used to estimate the sensitive volume. 
This volume multiplied by the duration of the analysis is the VT of the search. 
When evaluating the ratio of the VT in each analysis, we calculate the ratio of the VT by using multiple thresholds of the inverse false alarm rate (IFAR) that is assigned to each simulated signal in order to determine if a signal was detected by the search. 
We further present results for simulated signals with different chirp masses, $\mathcal{M} = (m_1 m_2)^{3/5}/(m_1 + m_2)^{1/5}$ for signals produced by the merger of objects with masses $m_1$ and $m_2$.
Errors for the ratio of the sensitive VT between analyses with and without DQ are estimated by calculating the VT ratio for 20 additional thresholds close to the chosen IFAR threshold. 

\begin{figure*}
  \centering
    \includegraphics[width=\textwidth]{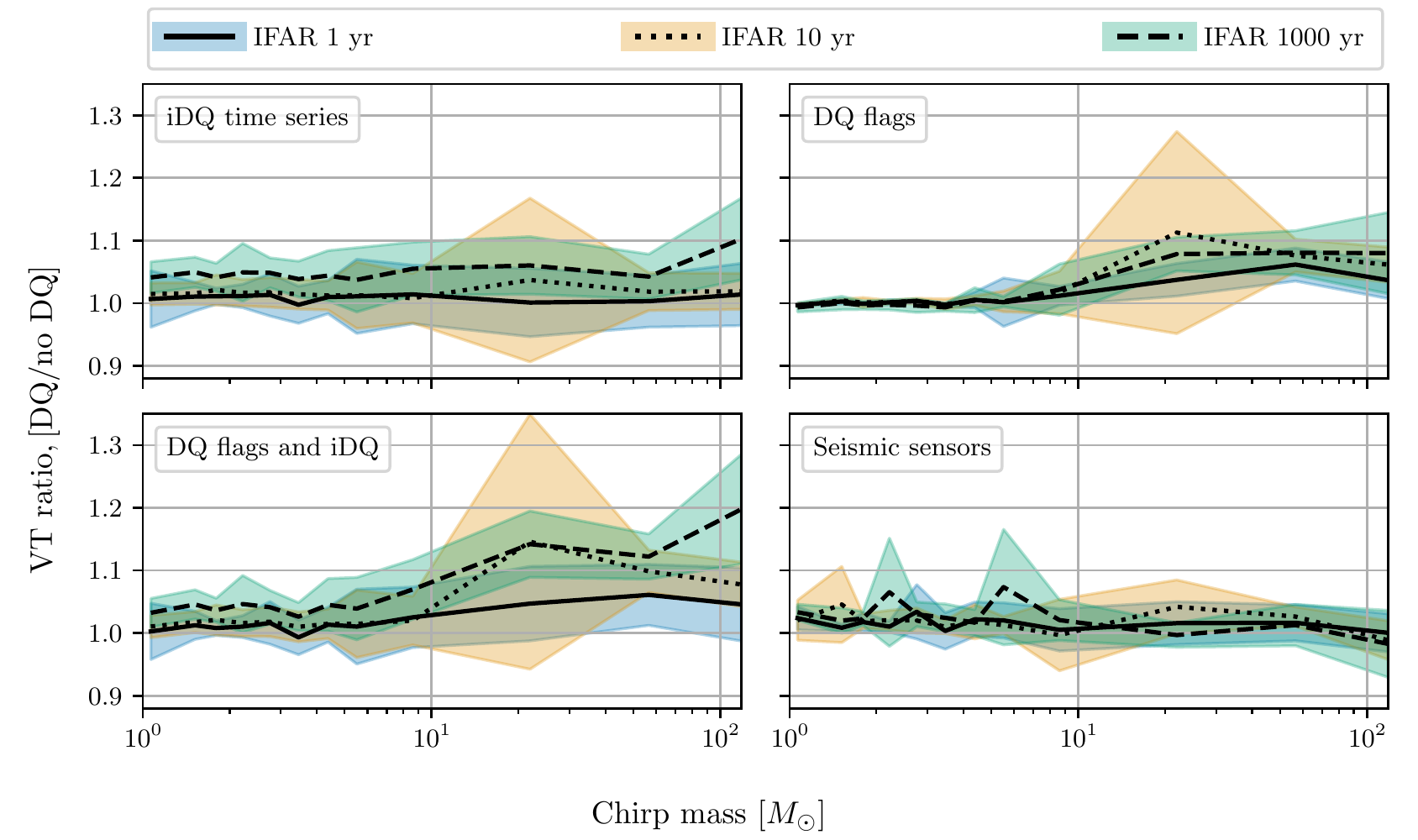}
\caption{The ratio of the sensitive volume-time (VT) for the \pycbc search, comparing the sensitivity of the search when using data quality (DQ) products as part of the ranking statistic versus using no data quality products. Three different detection thresholds are considered for each combination of data quality products in addition to a range of masses of simulated signals. Shaded regions correspond to the $1\sigma$ error 
in the measured VT ratio for each case. The measured VT ratio for all combinations of data quality products is above 1.0, indicating that the use of these data quality products only has a positive effect. Using both DQ flags and \idq yields the largest increase in VT of the 4 cases considered. }
\label{fig:vt}
\end{figure*}

\subsection{iDQ time series}
\label{sec:app_idq}

We analyze each of the chunks discussed above with \pycbc using the \idq log-likelihood time series produced in low-latency as a non-binary data quality stream.
The low-latency log-likelihood time series were produced by \idq using the OVL classifier~\cite{Essick:2013vga,godwin_diss}.
This classifier was trained using triggers from \result{844} LIGO auxiliary data streams at each detector.
This set includes all data streams that were determined to not be sensitive to gravitational-wave signals by LIGO detector characterization studies~\cite{LIGO:2021ppb}.
Separate instances of the classifier were trained for each interferometer used.
Each instance of the classifier was trained on triggers generated from \result{14 days} of detector data and used to make predictions until being replaced by a newly trained classifier.
After the training of each classifier completed, training of a new classifier began on the most recent 14 days of data.
We used the time series produced in low-latency, as opposed to time series produced offline at higher latency (and available at~\cite{idq_release}), because we found that the time series produced without the use of arbitrary chunk boundaries was a better predictor of data quality issues.

Before using the \idq time series in our analysis, we first pre-process the data stream.
We downsample the log-likelihood time series from \result{128 Hz} to \result{1 Hz}.
This is done by maximizing the \idq time series over
each integer second of data. 
The downsampled log-likelihood time series is then converted to percentiles, and each trigger is associated with the log-likelihood percentile at the time of the trigger.
Each template bin is divided into \result{200} sub-bins by the triggers' \idq log-likelihood percentiles, as described in Section~\ref{sec:nonbinary}.  

We find that including the \idq time series increases the sensitive VT of the search across the entire parameter space.
This increase in search sensitivity from using the \idq log-likelihood time series in \pycbc is shown in the upper left plot of Figure~\ref{fig:vt}.
We find that the gain in sensitivity generally increases with chirp mass, and is larger for higher choices of IFAR.
For triggers with chirp mass above \result{$80 M_\odot$}, we find a \result{$10\%$} increase in sensitive VT at an IFAR of \result{1000 years}.

Compared to the results of the Godwin et al. implementation of \idq into GstLAL~\cite{Godwin:2020weu}, our results show a larger increase in sensitive VT for the highest mass triggers. 
This is likely because we directly compute the time-dependent correction to the trigger rate instead of assuming an analytic formula for down-ranking triggers.
We find that the required correction during times corresponding to the highest percentiles of the \idq time series is lower than used in the Godwin et al. implementation.
It also does not down-rank any excess noise correlated with \idq time series percentiles below 50, even in cases where we include a correction.

\subsection{Alternate Binning Strategies}
\label{sec:app_bins}

In addition to the default method of binning the template parameter space that is explained in Section~\ref{sec:bins}, we investigate two alternate methods of binning the template parameter space.  

In the first alternate binning method, we construct six template bins with the bin edges in a geometric series between 0.15 seconds and 150 seconds. 
Thus the lowest bin edge is at 0.15 seconds, and each successive bin edge is larger by a factor of $\sqrt{10}$.
This method contains fewer total bins than the default method and the bins contain vastly differing numbers of triggers. 
However, the bin with the fewest total triggers was designed to contain roughly the same number of triggers as each of the bins in the default method. 

For the second alternate binning method, we first convert the trigger template durations into percentiles. 
We then construct five bins with bin edges placed at percentiles of $\{0, 6.25, 12.5, 25, 50, 100\}$ in the trigger template duration based on the entire bank of templates.
Similar to the first alternate method, the number of triggers in each bin is not the same, but the smallest bin is roughly the same size as each of the 10 bins when using the default method.
The only difference between this binning method and the default method is the location of the bin edges; this method places the bin edges in a geometric series with different amounts of triggers in each bin while the default method uses a linear series so that each bin contains the same number of triggers.  

\begin{figure}[tb]
  \centering
    \includegraphics[width=\columnwidth]{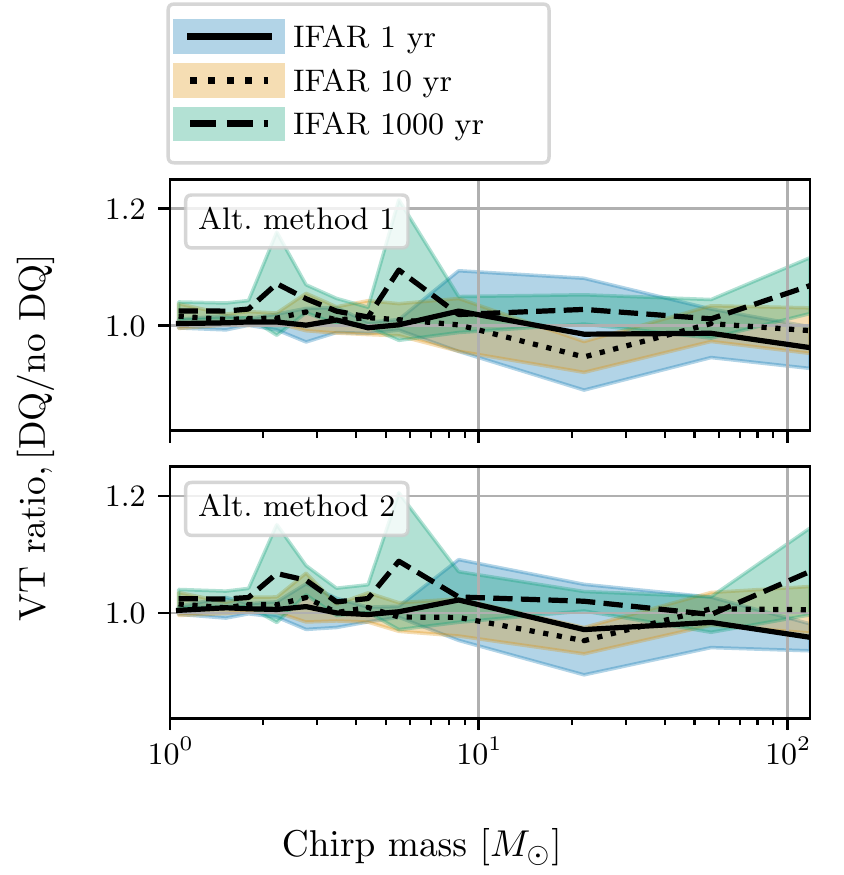}
\caption{The ratio of the sensitive volume-time (VT) for the \pycbc search when using the iDQ time series with different binning methods. In each case, the ratio of the measured VT when using the default binning method versus an alternate binning method is plotting. \textit{Top}: The ratio of the sensitive VT when using the default method of binning versus an alternate method that contains bins of different sizes. \textit{Bottom}: The ratio of the sensitive VT when using the default method of binning versus an alternate method that contains bins chosen based on the numerical value of the template durations. These alternate binning strategies perform very similarly, but the default method outperforms both alternate binning methods.}
\label{fig:bin_result}
\end{figure}

As part of evaluating which binning method to use in this work, we compared the sensitive VT for each binning method when using the \idq time series to analyze chunk 2.
The relative VT increases from using the default binning method as compared to each of the two  alternative binning strategies are shown in Figure~\ref{fig:bin_result}. 
These alternate binning strategies did not increase the sensitive VT as much as the default binning method did, so they were not used in any of our other analyses.

\subsection{Data quality flags}
\label{sec:app_flag}

\renewcommand{\arraystretch}{1.5}
\begin{table*}[tb]
\begin{tabularx}{\textwidth} { 
  | >{\raggedright\arraybackslash \hsize=0.35\hsize}X 
  | >{\raggedright\arraybackslash \hsize=0.58\hsize}X 
  | >{\raggedright\arraybackslash \hsize=2.12\hsize}X 
  | >{\raggedright\arraybackslash \hsize=0.35\hsize}X
  | >{\raggedright\arraybackslash \hsize=1.60\hsize}X | }
\hline
\textbf{Chunk} & \textbf{GPS interval} & \textbf{Data Quality Flag} & \textbf{Flag Time} & \textbf{Description} \\
\hline
1  & 1239641067 - 1240334090 &  
L1:DCH-PEM\_EY\_ACC\_BEAMTUBE\_OMICRON\_GT\_100 & $1.06 \%$ & 70 Hz periodic glitches due to an automated camera shutter in the End-Y station at LIGO Livingston. \\
\hline
2  & 1241724868 - 1242485150 &  H1:DCH-EARTHQUAKE\_CS\_Z\_BLRMS\_GT\_1000 & $ 0.48 \%$ & Non-stationary noise due to high ground motion at LIGO Hanford. \\
  &  &  L1:DCH-THUNDER\_MIC\_BP\_GT\_300 & $ 0.05 \%$ & Excess noise due to thunderstorm at LIGO Livingston. \\
\hline
3 & 1262192836 - 1262946499 &  L1:DCH-WHISTLES & $ 0.45 \%$ & Glitches caused by radio frequency (RF) beat notes at LIGO Livingston.  \\
\hline
4 & 1263751734 - 1264528232 &  L1:DCH-WHITENED\_RF45\_AM\_CTRL\_GT\_1P75 & $ 0.27 \%$ & Glitches due to 45 MHz control signal at LIGO Livingston.  \\
  &  &  L1:DCH-WHISTLES & $ 0.83 \%$ & Same as chunk 3.  \\
\hline
5 & 1264528056 - 1265133171 &  L1:DCH-WHITENED\_RF45\_AM\_CTRL\_GT\_1P75 & $ 0.54 \%$ & Same as chunk 4. \\
   &  &  L1:DCH-WHISTLES & $ 0.16 \%$ & Same as chunk 3. \\
\hline
\end{tabularx}
\caption{
A list of the time periods analyzed and the data quality flags used as data quality streams in this work. All data quality flag names are sourced from~\cite{T2100045}. Flag time refers to the analyzable time impacted by each individual flag in the analyses period.}
\label{tab:flag} 
\end{table*}

\begin{figure}[tb]
  \centering
    \includegraphics[width=\columnwidth]{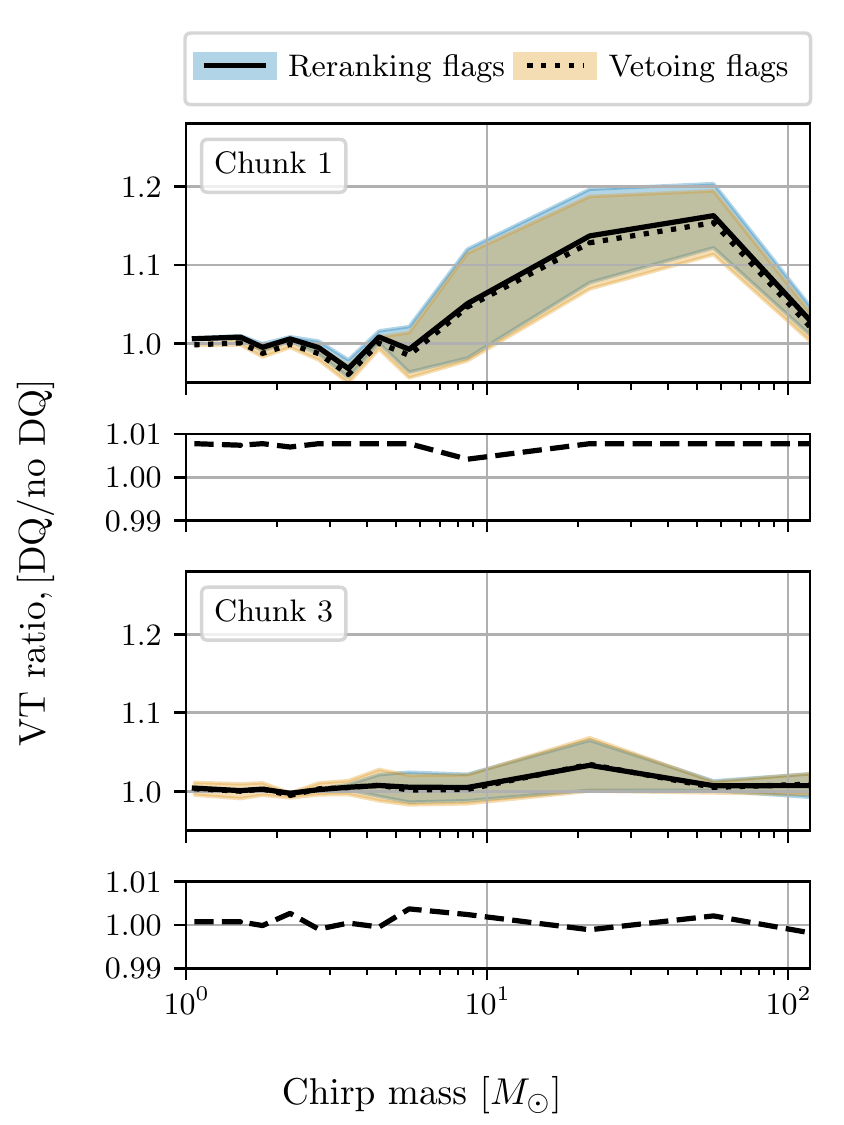}
\caption{The ratio of the sensitive volume-time (VT) for the \pycbc search when using data quality (DQ) flags to rerank \pycbc candidates (blue) or vetoing candidates (orange). The sensitivity is calculated at fixed inverse false alarm rate of 10 years. \textit{Top}: Increase in search sensitivity for chunk 1, an analysis where a data quality flag was known to have a positive effect. The ratio of the search sensitivity when using reranking versus vetoing candidates is shown in the second panel. \textit{Bottom}: Increase in search sensitivity for chunk 3, an analysis where a data quality flag was known to have minimal effect. The ratio of the search sensitivity when using reranking versus vetoing candidates is shown in the fourth panel. In both cases, reranking times during data quality flags only increases the sensitivity of the search compared to vetoing. 
}
\label{fig:flag_result}
\end{figure}

We next investigate the benefits of using data quality flags as a part of our time-dependent noise model. 
The LIGO and Virgo collaboration uses a wide variety of data quality flags to indicate periods when environmental or instrumental noise sources are likely to affect the quality of the strain data. 
Currently, the \pycbc search uses these flags to remove triggers during these time periods from the analysis.
However this reduces the analyzable time and could cause the search to miss some gravitational-wave signals.
We can instead use these data quality flags as binary data streams to 
 take into account the expected increase in the trigger rate and re-weight the detection statistic of triggers accordingly. 
Table \ref{tab:flag} details the data quality flags active during the analysis periods that we chose to analyze. 
These data quality flags are released via GWOSC as a single, combined data stream~\cite{LIGOScientific:2019lzm} but are not currently publicly available separately.
We chose to consider these data quality flags as multiple data streams, as each data quality flags was designed to target a different noise source, making it easier to measure the effect of these noise sources on the \pycbc trigger rate. 

We also choose to calculate the time-independent portion of the \pycbc noise model
after removing candidates that are present during the data quality flag segments. 
These candidates are still considered potential astrophysical candidates and their significance is estimated as described in section \ref{sec:binary}.
We find that excluding these time periods when calculating the time-independent terms in the \pycbc noise model increases the sensitivity as compared to including them. 

The upper right panel of Figure \ref{fig:vt} shows that including data quality information in the \pycbc search increases its sensitivity, in particular for high mass binaries. 
In this region of the parameter space, the number of detectable gravitational-wave signals increase by \result{10\%}.
However, the sensitivity gains vary greatly between analysis periods. 
As shown in Figure \ref{fig:flag_result}, including data quality information in the search of chunk 1 data increases the sensitivity to signals from binary black hole mergers up to ~15\%. 
On the other hand, our approach has just a small effect for chunk 3. 
In fact, this period is dominated by glitches that are effectively identified and down-ranked by the \pycbc consistency tests \cite{Nitz:2017lco}. 

Figure \ref{fig:flag_result} also shows how our approach compares to the previous method that \pycbc use to incorporate data quality flags, namely using the data quality flag segments to veto candidates.
The improvements in sensitivity are due to the increased analyzable time. 
This increase in sensitivity compared to using data quality flags as vetoes directly translates into more events that can be detected by \pycbc. 
Although the amount of time vetoed by data quality flags in recent observing runs is less than \result{$1\%$}~\cite{LIGO:2021ppb}, 
the high rate of detections makes it likely that some events would be missed or recovered with less significance by chance due to vetoes. 

One such event, GW$200129\_065458$, was identified by the ``PyCBC-Broad'' search in GWTC-3 as a coincident signal between LIGO Hanford and Virgo~\cite{GWTC-3}. 
This event was not identified as a three-detector coincidence because the related trigger at LIGO Livingston was vetoed by a data quality flag. 
We find that using data quality flags for reranking triggers instead of vetoing them allows this event to also be identified at LIGO Livingston with high significance.

\subsection{Multiple data quality products}
\label{sec:app_combined}

In addition to considering the use of the \idq time series and data quality flags separately, we investigated the benefit of using both types of data quality products at once. 
In cases when a trigger is down-ranked by both the \idq time series and a data quality flag, only the larger amount of down-ranking was used, as described in section~\ref{sec:multi}.
We also include data quality flag information in the same way as in the previous section; triggers during data quality flags are removed when the time-independent noise model is calculated but included when candidates are identified. 

We find that including both the \idq time series and data quality flags increases the sensitivity of \pycbc by \result{20\%} compared to 
no use of data quality products, as shown in the lower left panel of Figure \ref{fig:vt}.
This is roughly in line with what would be expected from adding the sensitivity increases from the individual data quality flag and \idq results.
Although the auxiliary data streams that were were used to create the considered data quality flags are also used by \idq, this result suggests that the data quality issues identified by each product are distinct. 

\subsection{Seismic monitors}
\label{sec:app_eq}

Seismic activity is a major source of noise for LIGO~\cite{Coughlin:2014wfa,SeisVeto,LIGO:2020pzq}. 
Seismic noise can couple into the detector and appear as scattered light glitches~\cite{Accadia:2010zzb,Soni:2020rbu}.
We use seismic trend data as another example of a non-binary data quality stream. 

For the input data quality stream for our analysis, we use accelerometer data from the corner station at each observatory. 
These monitors measure ground motion in the direction perpendicular to the arms of the interferometer. 
The chosen data streams are focused on ground motion 
from 0.03 -- 0.1 Hz, 
which is often referred to as the ``earthquake band''
as earthquakes are the main contributor to ground 
motion at these frequencies.

For this investigation, we choose a single 
analysis period, chunk 2, covering from \result{5 May 2019} to \result{21 May 2019}.
Similar to the previous investigations, 
this time was chosen due to the known presence of a data quality issue
that could be correlated with this data stream. 
This seismic data is not available for public use via GWOSC, 
but is displayed on the public ``Detector Status'' pages~\cite{gwosc_status}.

We found that this increased the sensitivity of the search by as much as \result{5\%} in some regions of the trigger parameter space. 
The increase in sensitivity from using these seismic sensors across different template masses is shown in the lower right panel of Figure~\ref{fig:vt}.
For most of the parameter space, only a marginal increase in sensitivity is measured. 
Incorporating additional sensor data may further increase these sensitivity gains. 
As the methods presented in this work are fully generic for any time series, 
any useful auxiliary information can be further incorporated into the search.

\subsection{Large numbers of auxiliary monitors}
\label{sec:app_gwosc}

In each observation run, hundreds of thousands of auxiliary data streams are recorded for the full duration of the run and could potentially be incorporated into the 
\pycbc search using the methods described in this work. 
However, at the time of publication, the LIGO Scientific Collaboration has only publicly released 
auxiliary data streams for a single data segment for a small subset of streams. 
This data release, containing data from \result{1169} data streams for
3 hours around GW170814~\cite{GW170814} is available at~\cite{auxdata}.
Although this amount of data is not sufficient to test if these data streams can be used to increase the sensitivity of the \pycbc search, we use this data release to demonstrate how this method can be applied for 
a large number of separate auxiliary monitors.
We choose to only include an auxiliary data stream if a data stream with the same name was available from both sites.
This reduced the total number of data streams used to \result{1126}.

For this investigation, we made multiple changes to the standard workflow to both increase the likelihood that relevant features of included auxiliary data streams are identified as correlated with the \pycbc trigger rate and decrease the computational cost. 
When possible, the auxiliary data was bandpassed and the root-mean-square (RMS) of the data was calculated with a \result{1 second} stride. 
Data streams with a sample rate of higher than \result{100 Hz} were bandpassed between \result{10 Hz} and \result{100 Hz} while streams with sample rates between \result{10 and 100 Hz} were bandpassed between \result{1 Hz} and \result{10 Hz}. 
Data streams with sample rates below \result{10 Hz} were instead set to the maximum value of that stream in each \result{1 second} stride. 
As this investigation was not used to estimate the sensitivity of the search, we used a small template bank targeting chirp masses between \result{10$M_\odot$} and \result{40$M_\odot$}. 
Due to this smaller template bank and the small amount of data considered, we lowered the SNR threshold used to calculate the trigger rate to \result{4.5} in order to increase the number of triggers considered.

Due to the large number of auxiliary data streams considered,
it is highly likely that some sources of noise will be observed by multiple streams. 
In this case, the down-ranking applied is given by the description in Section~\ref{sec:multi}, 
namely that the maximum measured down-ranking among all data streams will be applied.

\begin{figure}
  \centering
    \includegraphics[width=\columnwidth]{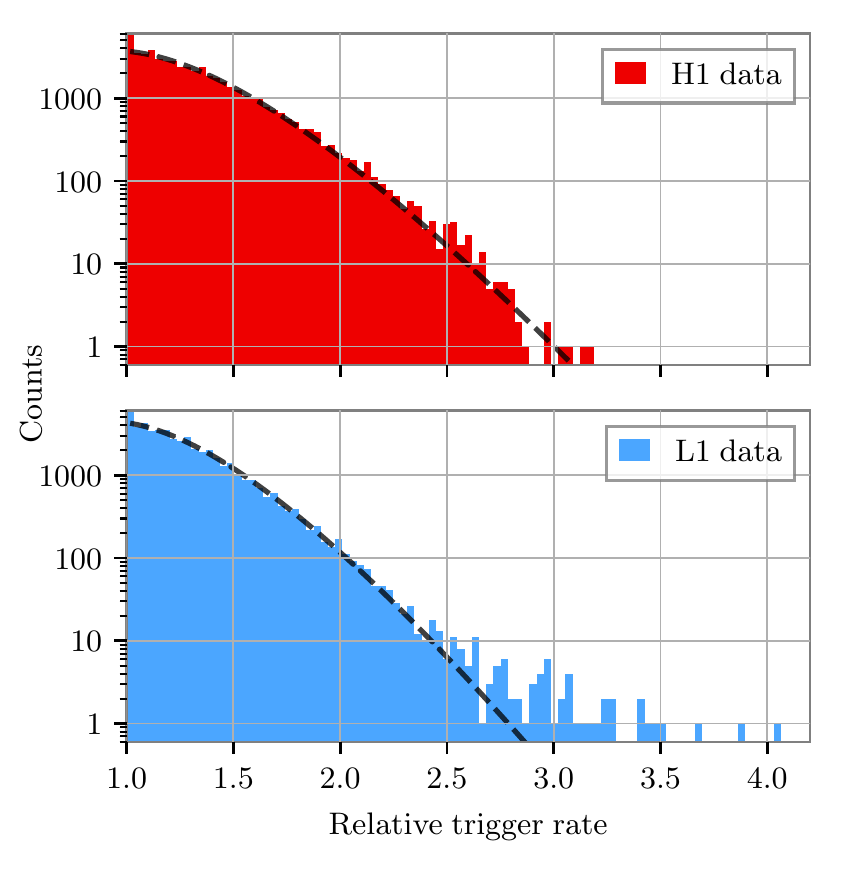}
\caption{Histograms of the measured trigger rate in each data quality bin from the \result{1126} auxiliary data streams considered in this analysis. The relative trigger rate is the ratio of the rate of triggers in each data quality bin versus the average rate of triggers at each detector. The data is fit to a Poisson distribution, shown as a black dotted line. Data from LIGO Hanford (top) shows no clear outliers, while data from LIGO Livingston (bottom) includes a small numbers of outliers based on the fitted distribution.}
\label{fig:aux_result}
\end{figure}

The measured correlation between trigger rate in \pycbc
and the \result{1126} data streams is shown in the upper panel of Figure~\ref{fig:aux_result}. 
If the data streams were uncorrelated with the rate of triggers, we would expect that the distribution of the measured trigger rate in each bin would follow a Poisson distribution. 
A fit of the data with this distribution is shown in Figure~\ref{fig:aux_result} as a black dotted line.
For most of these auxiliary data streams, there is no clear correlation observed between the
data stream and the rate of triggers in \pycbc. 
However, for \result{10} data streams at LIGO Livingston, there is at least one data quality bin 
with a measured relative trigger rate above \result{3.25}. 
This threshold is much higher than expected due to chance based on the fitted Poisson distribution.  

The data streams that show the strongest correlation with the \pycbc trigger rate includes sensors designed to detect ground motion and magnetic noise at LIGO Livingston. 
Monitors of ground motion~\cite{SeisVeto, Soni:2020rbu} and magnetic noise~\cite{Nuttall:2018xhi} are known to be correlated with glitches in LIGO data,
so it is not surprising that these data streams are the most significant outliers in the small amount data considered in this investigation. 
Details about these \result{10} outliers are included in Table~\ref{tab:chans}.

\renewcommand{\arraystretch}{1.5}
\begin{table*}[tb]
\begin{tabularx}{0.9\textwidth} { 
  | >{\raggedright\arraybackslash \hsize=1.25\hsize}X 
  | >{\raggedright\arraybackslash \hsize=1.30\hsize}X
  | >{\raggedright\arraybackslash \hsize=0.45\hsize}X | }
\hline
\textbf{Data stream name} & \textbf{Data stream description} & \textbf{Maximum trigger rate} \\
\hline
L1:PEM-CS\_ACC\_LVEAFLOOR\_BS\_Z\_DQ & LVEA Accelerometer & 4.05 \\
L1:HPI-BS\_BLND\_L4C\_RX\_IN1\_DQ & Pre-isolator motion in the global ifo basis & 3.87 \\
L1:HPI-ITMY\_BLND\_L4C\_RX\_IN1\_DQ & Pre-isolator motion in the global ifo basis & 3.66 \\
L1:HPI-HAM3\_BLND\_L4C\_RX\_IN1\_DQ & Pre-isolator motion in the global ifo basis & 3.51 \\
L1:HPI-HAM3\_BLND\_L4C\_VP\_IN1\_DQ & Pre-isolator motion in the global ifo basis & 3.49 \\
L1:PEM-EX\_MAG\_VEA\_FLOOR\_Y\_DQ & Magnetometer near ETMX chamber & 3.43 \\
L1:HPI-ITMX\_BLND\_L4C\_RX\_IN1\_DQ & Pre-isolator motion in the global ifo basis & 3.42 \\
L1:ASC-INP1\_P\_IN1\_DQ & Error signal for input beam in pitch & 3.42 \\
L1:HPI-ITMX\_BLND\_L4C\_RY\_IN1\_DQ & Pre-isolator motion in the global ifo basis & 3.27 \\
L1:PEM-CS\_ACC\_IOT1\_IMC\_Z\_DQ & LVEA Accelerometer & 3.26 \\
\hline
\end{tabularx}
\caption{
List of the auxiliary data streams used in the search of 3 hours of data around GW170814
that are highly correlated with the rate of \pycbc triggers.
All data streams with a maximum trigger rate of over \result{3.25} are listed. 
Descriptions of each data stream are sourced from~\cite{auxdata}. 
}
\label{tab:chans} 
\end{table*}

As auxiliary data is only available for 3 hours, 
we were not able to use this improved noise model
to reanalyze the full LIGO data set and identify 
new gravitational-wave candidates. 
However, if auxiliary data does become available, 
this method would allow this data to be directly 
used in searches for gravitational waves. 

\section{Conclusions}
\label{sec:conc}

We have demonstrated
a novel method of directly using auxiliary data
in a search for gravitational waves.
This method can be applied to both the original auxiliary data and derived data quality products
that are distributed alongside the strain data. 
Although this method was applied to the 
\pycbc search for compact binaries, 
similar methods can be incorporated to other
search algorithms for both 
compact binaries~\cite{Sachdev:2019vvd,Aubin:2020goo,Chu:2020pjv,Venumadhav:2019tad}
and other gravitational-wave sources~\cite{Klimenko:2015ypf, Lynch:2015yin}. 

With currently available data quality products, 
this method was able to increase the sensitivity 
of the \pycbc search across a 
wide range of masses. 
We find that the number of detectable gravitational-wave events 
is increased by up to \result{20\%} for a subset of the gravitational-wave signal population
when using a combination of data products.
This method also removes the need for data quality products
to be curated before use by \pycbc, 
reducing the time required to fully analyze LIGO data. 

Ultimately, the benefits of this method
are limited by the available data quality streams. 
Using data quality streams that are highly predictive
of a high rate of \pycbc triggers will naturally
increase the benefits of this method. 
However, compared to previous methods of incorporating 
data quality information, 
the method outlined in this work
will not decrease the overall sensitivity if
the auxiliary data stream is uninformative.  

The versatility of this method will reduce the required
effort of LIGO data quality experts to 
produce derived data quality products. 
Rather than using hand-tuned binary data quality flags, 
this method allows the \pycbc search to directly 
ingest the relevant auxiliary data stream. 
In addition, directly ingesting 
the auxiliary data stream may be more beneficial 
to the overall sensitivity of the search. 

Similar methods can be applied to the low-latency version of
the \pycbc search, \texttt{PyCBC Live}~\cite{Nitz:2018rgo,DalCanton:2020vpm}.
One practical difference for a low-latency implementation of this method
is that most auxiliary data streams 
are not available at the latencies required 
for detection. 
At present, only a subset of data quality flags 
and the \idq time series are available at the required latency. 

There are a number of areas of improvement for this method
that could be explored in future works. 
First, 
we could add additional time dependence to our improved noise model. 
This method does not  account for variance in the $\alpha(t)$ parameter, 
which also could impact sensitivity of the search. 
There is also an assumption that the auxiliary data stream 
does not include any time delay between the auxiliary data and the time of the
\pycbc trigger. 
This may not be valid for low mass signals
that last many seconds or minutes. 
Finally, this method could be improved by better addressing the case of multiple 
correlated input data streams. 

This work presents a novel method that is able to 
directly use the large datasets produced at a gravitational-wave
observatory in an astrophysical analysis. 
At present, this data is not publicly available. 
Hence, the maximum benefits of this work can only be realized
by internal LIGO analyses. 
However, this method demonstrates one such practical use
of directly using this dataset in astrophysical analyses
and provides additional motivation for their curation and release.

\acknowledgments
The authors thank the LIGO-Virgo-KAGRA PyCBC and Detector Characterization groups for their input and suggestions during the development of this work. 
We would like to thank Patrick Godwin for productive discussions on how to best utilize iDQ time series data, and to Gareth Cabourn Davies for their comments during internal review of this paper. 
DD is supported by the NSF as a part of the LIGO Laboratory. 
MT is supported by the NSF through grant PHY-2012159.
SM is supported by a STFC studentship.
LKN thanks the UKRI Future Leaders Fellowship for support through the grant MR/T01881X/1.

This material is based upon work supported by NSF’s LIGO Laboratory 
which is a major facility fully funded by the 
National Science Foundation.
LIGO was constructed by the California Institute of Technology 
and Massachusetts Institute of Technology with funding from 
the National Science Foundation, 
and operates under cooperative agreement PHY-1764464. 
Advanced LIGO was built under award PHY-0823459.
The authors are grateful for computational resources provided by the 
LIGO Laboratory and supported by 
National Science Foundation Grants PHY-0757058 and PHY-0823459.
This work carries LIGO document number P2200078.

\bibliography{pycbc_dq.bbl}

\end{document}